\documentclass{aa}

\usepackage[varg]{txfonts}
\usepackage{natbib}
\usepackage{graphicx}
\usepackage{placeins}

\newcommand{\kms}{\ensuremath{\mathrm{km}\,\mathrm{s}^{-1}}}
\newcommand{\radyn}{{\textrm{RADYN}}}

\begin{document}

\title{The F-CHROMA grid of 1D RADYN flare models}
\author{Mats Carlsson\inst{1,2}
\and Lyndsay Fletcher\inst{3,1,2}
\and Joel Allred\inst{4}
\and Petr Heinzel\inst{5,6}
\and Jana {Ka{\v s}parov{\'a}}\inst{5}
\and Adam Kowalski\inst{7,8,9}
\and Mihalis Mathioudakis\inst{10}
\and Aaron Reid\inst{10}
\and Paulo J. A. Sim\~oes \inst{11,3}
}

\institute{
Institute of Theoretical Astrophysics, University of Oslo, P.O. Box
1029 Blindern, N-0315 Oslo, Norway
\and
Rosseland Centre for Solar Physics, University of Oslo, P.O. Box
1029 Blindern, N-0315 Oslo, Norway
\and
SUPA School of Physics and Astronomy, University of Glasgow, Glasgow G12 8QQ, UK.
\and
NASA/Goddard Space Flight Center, Code 671, Greenbelt, MD 20771, USA
\and
Astronomical Institute of the Czech Academy of Sciences, 25165 Ond\v{r}ejov, Czech Republic
\and
University of Wroc{\l}aw, Center of Scientific Excellence - Solar and Stellar Activity, Kopernika 11, 51-622 Wroc{\l}aw, Poland
\and 
National Solar Observatory, University of Colorado Boulder, 3665 Discovery Drive, Boulder, CO 80303, USA
\and
Department of Astrophysical and Planetary Sciences, University of Colorado, Boulder, 2000 Colorado Ave, CO 80305, USA
\and
Laboratory for Atmospheric and Space Physics, University of Colorado Boulder, 3665 Discovery Drive, Boulder, CO 80303, USA
\and
Astrophysics Research Centre, School of Mathematics and Physics, Queen’s University Belfast, BT7 1NN, Northern Ireland, UK
\and
Center for Radio Astronomy and Astrophysics Mackenzie, Engineering School, Mackenzie Presbyterian University, S\~ao Paulo, Brazil
}

\date{\today / -  }

\authorrunning{Carlsson et al.}
\titlerunning{F-CHROMA grid}
\abstract
{Solar flares are the result of the sudden release of magnetic energy
  in the corona. Much of this energy goes into accelerating charged particles
  to high velocity. These particles travel along the magnetic field  and the energy is dissipated when the density gets high enough,
  primarily in the solar chromosphere.
Modelling this region is difficult because the radiation energy
balance is dominated by strong, optically thick spectral lines.} 
{Our aim is to provide the community with realistic simulations of a
  flaring loop with an emphasis on the detailed treatment of the
  chromospheric energy balance. 
This will enable a detailed
comparison of existing and upcoming observations with synthetic observables from the simulations, 
thereby elucidating the complex interactions in a flaring chromosphere.} 
{We used the 1D radiation hydrodynamics code \radyn\ to perform
  simulations of the effect of a beam of electrons injected at the apex
  of a solar coronal loop. A grid of models was produced, varying the
  total energy input, the steepness, and low-energy cutoff of the beam energy spectrum.} 
{The full simulation results for a grid of models are made available online. 
Some general properties of the simulations are discussed.} 
{} 

\maketitle
 
\section{Introduction}\label{sec:introduction}
Interpretation of the radiation produced by the flaring solar chromosphere needs to be underpinned by the numerical modelling of the complex and interlinked physics of this region. In a flare, the atmosphere is strongly disturbed and non-equilibrium physics is critical, so this is particularly challenging. So far, this has restricted us, in practical terms, to 1D approaches in which the equations of hydrodynamics, radiation transfer, and also the atomic rate equations for the energetically most important lines and continua are solved simultaneously. We focus in this paper on the \radyn\ code, though there are other codes capable of tackling this problem at different levels of sophistication, for example FLARIX \citep{2009A&A...499..923K,2010ITPS...38.2249V,2016IAUS..320..233H}, HYDRAD \citep{2003A&A...401..699B,2013ApJ...770...12B}, and HYDRO2GEN \citep{2019A&A...623A..20D}. 

The chromosphere responds in dramatically different ways depending on the character of the flare energy input (e.g. the energy input rate and the column depth at which the energy deposition primarily occurs). The resulting interplay between temperature and density variations, and bulk flows caused by plasma expansion determines the detailed line profiles of the common spectral lines used to diagnose chromospheric properties. These spectral lines can be used to deduce a flare model chromosphere, using traditional semi-empirical approaches of iterating models towards a best fit to data \citep[e.g.][]{1980ApJ...242..336M}. Alternatively, models can be run that are `tuned' to data, so as to match inputs (electron beam parameters), outputs (e.g. observed intensities or line profiles), or both \citep[e.g.][]{2015ApJ...813..125K,2017ApJ...836...12K,2017A&A...605A.125S}.

However, running such models is quite a time-consuming and specialised task, so to facilitate the comparison of observations with models, and also to promote a broader understanding of the physics of the flare chromosphere, we have created a database of flare atmospheres modelled with the \radyn\ code. In the version of \radyn\ used (see below for a full description), it is assumed that the flare energy input is in the form of a beam of non-thermal electrons that are accelerated in the corona  with a power-law distribution, and characterised by their total energy per area, spectral index, and minimum electron energy (`low-energy cutoff'). The simulations are run for a range of different values of beam total energy, spectral index, and low-energy cutoff, providing a description of the expected evolution of a 1D column of chromosphere, at 0.1\,s intervals, in large and small flares. This database was an output of the Flare Chromospheres: Observations, Models and Archives (F-CHROMA) collaboration supported by the Seventh Framework Programme of the European Union,
and is known as the \radyn\ F-CHROMA database, 

The \radyn\ code calculates time-evolving atmospheric parameters from photosphere to corona (e.g. temperature, number densities, level populations, and flow speed) and emission in a number of strong chromospheric spectral lines and continua. Models from the \radyn\ F-CHROMA database have already been used for several different purposes, with the atmospheric parameters often used as input to other packages such as RH \citep{2001ApJ...557..389U}, Lightweaver \citep{2021ApJ...917...14O}, and CHIANTI \citep[e.g.][]{1997A&AS..125..149D}. This is done to evaluate the emission in lines  and continua not directly calculated by \radyn, thus expanding the range of problems that can be studied.

Using the whole \radyn\ F-CHROMA grid of models, or substantial fractions of it, one can explore the response of commonly used observables to flare energy input with different properties such as total flux or the electron beam spectral index. \cite{2020ApJ...893...24S} used the grid to study the effects on the \ion{Fe}{I} line at 6173\,\AA,\, which is formed in the deep atmosphere and used in measurements of the photospheric magnetic field, and \cite{2021ApJ...915...16M} looked at predicted photospheric velocity signals during flares from this and two other \ion{Fe}{I} lines.  \cite{2020ApJ...894L..21R} evaluate the effect of flare-produced  hot `bubbles' in the upper chromosphere on the profiles of the H$\alpha$ and \ion{Ca}{II} 8542\,\AA\ lines. \cite{2019ApJ...871....2S} used 20 grid models to examine theoretical correlations between energy deposition and Doppler shifts deduced from \ion{C}{II} and \ion{Fe}{XXI} profiles, both of which are observed by the Interface Region Imaging Spectrograph (IRIS) \citep{2014SoPh..289.2733D}. \cite{2020Ge&Ae..60.1038M} used atmospheric parameters from the grid to predict the sub-terahertz emission from a range of flare-heated plasma.

Individual F-CHROMA models matching the inferred energy input parameters of specific flares have also been selected for particular observational studies, for example, 
\citet{2017ApJ...850...36C} used the grid to investigate the formation and evolution of H$\alpha$ and H$\beta$ lines during a solar flare in a unique case of simultaneous observations of both lines by the Interferometric Bidimensional Spectrometer \citep[IBIS,][]{2006SoPh..236..415C} and the Rapid Oscillations in the Solar Atmosphere \citep[ROSA,][]{2010SoPh..261..363J}, respectively.
\cite{2018SciA....4.2794J} used the atmospheric properties of a  single weak-flare model to calculate the optical depth in \ion{Si}{IV} in the interpretation of IRIS observations, while \cite{2020ApJ...897L...6H} compare observations and simulations of the enhanced absorption in the \ion{He}{I} 10830~{\AA} line in the early phase of a solar flare. The sub-teraherz emission has also been calculated for a specific model matching the parameters of the event of July 4, 2012 \citep{2021Ge&Ae..61.1045M}. 

The F-CHROMA grid has also been used to develop and test new methods. \cite{2019ApJ...871...18R} tested a new technique for a fast calculation of out of Local Thermodynamic Equilibrium (non-LTE) effects in a flare against a single model; while the entire grid of atmospheres, and their corresponding H$\alpha$ and Ca~\textsc{II} 8542~\AA~line profiles, have been used by \cite{2019ApJ...873..128O} to train an invertible neural network, making it possible to generate a RADYN-like atmosphere from observed spectral line profiles. 

The purpose of this paper is to describe the F-CHROMA RADYN grid, and also to  encourage its exploitation and exploration. 
This will be valuable for detailed comparison with individual flares where enough information about the energy input is available to be able to identify the closest model match. 
In the grid, the energy input is described by several electron beam parameters. The range of beam parameters adopted in
our grid was defined from the parameters deduced from decades of hard X-ray observations
and statistical studies of solar flares \citep[e.g.][]{2011SSRv..159..263H, 2013A&A...552A..86W, 2016ApJ...832...27A, 2019ApJ...881....1A}.
However, each set of parameters lead to different atmospheric behaviour, making the 
response to the beam heating a complex problem. Therefore, the grid of models should also be considered as a series of numerical experiments, which can be used to develop an understanding of the relationships between observables and hidden atmospheric properties across a range of flares.

\section{RADYN}\label{sec:radyn}
The simulations described here have been performed with the
1D radiation hydrodynamics code,
\radyn,\ developed by \citet{1992ApJ...397L..59C,1995ApJ...440L..29C,1997ApJ...481..500C,2002ApJ...572..626C} with flare
physics added by 
\citet{1999ApJ...521..906A} and \citet{2005ApJ...630..573A, 2015ApJ...809..104A}. 
 \radyn\ solves the coupled equations of conservation of mass, momentum, energy, charge, and population rate
equations on an adaptive grid; readers can refer to the above references for details. 
Our version of the code
is essentially that of \citet{2015ApJ...809..104A}, and we refer the reader to that source for a more detailed description of the beam
physics. There are differences in the initial atmosphere and the model atoms employed as described below.
Additional heating from return currents is not included in the models presented here.

The optically thin radiative losses are calculated summing up all contributions, except those considered in the detailed radiative transfer, using the CHIANTI v.7 database \citep{1997A&AS..125..149D, 2012ApJ...744...99L}. 
Half of this radiation escapes the atmosphere outwards, but half is directed inwards and is ultimately absorbed by the underlying, denser atmosphere. 
This backwarming is taken into account by integrating the emissivities from the coronal and transition region part of the loop for each wavelength in the photoionisation continua included in the detailed radiative transfer (see table~\ref{tab:trans}). 
This incoming intensity is then included as a boundary condition for the radiative transfer calculation. Readers can refer to \citet{2015ApJ...809..104A} for more details.

Background opacity from elements other than those treated in detail have been included assuming LTE, with photoionisation cross-sections from  The Opacity Project dataBASE (TOPBASE\footnote{http://cdsarc.u-strasbg.fr}).
We have included nonthermal collisional rates due to beam electrons for the hydrogen 1-c (continuum), 1-2, 1-3, and 1-4 transitions following \cite{1993A&A...274..917F}. We used the method of \citet{1981JQSRT..26..329Y} and \citet{1985A&AS...60..425A} to model the nonthermal collision rates for neutral and singly ionised helium.
Readers can refer to \citet{2015ApJ...809..104A} for more details.  
The version of RADYN used for the F-CHROMA grid is openly available\footnote{
https://folk.universitetetioslo.no/matsc/radyn/radyn\_fchroma.tar}
with all input files with references to the sources of the atomic data included in the atomic data files.

\section{Model atoms}\label{sec:atoms}

We have included model atoms for hydrogen, singly ionised calcium,  and helium (see tables~\ref{tab:h}-\ref{tab:he}). 
We set the calcium abundance to 6.34 \citep{2009ARA&A..47..481A} and the helium abundance to 11 on the usual logarithmic scale where the abundance of hydrogen is 12.
We note that some helium levels 
have been treated as terms instead of individual levels (\ion{He}{i} $1s\,\,2p\,\,^{3}\!P^{o}$ and \ion{He}{ii} $2p\,\,^{2}\!P^{o}$). In total, we have included 
41 transitions that are numbered consecutively in the output files, as detailed in table~\ref{tab:trans}. One of the transitions (\# 15) has been included as 
a continuum transition in the hydrogen model atom, but with a zero photoionisation cross-section giving zero contribution to the radiative
rates. This transition has only been  included in order to account for the energy balance due to photons longwards of the Pfundt continuum (wavelengths between 22386.68 and 40000~\AA). All bound-bound transitions have been treated with 201 frequency points and the bound-free transitions have been 
covered with 4-34 frequency points, giving a total of 4795 frequency points.

All bound-bound transitions have been treated assuming complete frequency redistribution (CRD). This is not a good approximation for the hydrogen Lyman lines or the calcium H \& K lines where partial frequency redistribution (PRD) effects are important \citep[e.g.][]{1981ApJS...45..635V}.
Two recipes have been used in the literature to mimic effects of PRD in the energy balance in CRD calculations of hydrogen: (i) truncation of Voigt profiles at around six Doppler widths from the line centre and (ii) using Doppler profiles instead of Voigt profiles (see \citet{2012ApJ...749..136L} for a discussion). 
With \radyn ,\ the former approximation has been used until now, but the disadvantage is that for large velocities in the atmosphere (of the order of 50 \kms), the full absorption profile is shifted out of the $\pm$ 6 Doppler width passband giving errors in the energy balance. 
Such large velocities are common in flare simulations, and we have therefore opted for the second approximation for the grid presented here. 
We have thus neglected radiative damping and van der Waals broadening, but we have included Stark broadening in our profile calculation for the hydrogen Lyman lines. 
These approximate treatments of PRD effects for hydrogen Lyman lines have been developed for quiet Sun applications and further study is really needed to find the best treatment for flare simulations.

Using CRD for the \ion{Ca}{ii} H \& K lines typically produces a radiative cooling a factor of two larger than if the proper PRD treatment is used \citep{2002ApJ...565.1312U}.
On the other hand, the \ion{Mg}{ii} h \& k lines give a similar amount of cooling as the \ion{Ca}{ii} H \& K lines. By not including Mg in the modelling, we have compensated for the too large cooling by the CRD treatment of \ion{Ca}{ii} lines. 
Again, these relations have only been tested for quiet Sun conditions and an assessment should be done for flaring conditions as well.

\begin{table}[htpb]
\caption{Hydrogen energy levels.}
\label{tab:h}
\begin{tabular}{rrrl}
Level & E (cm$^{-1}$) & E (eV) & Designation\\
\hline
0 & 0.000 & 0.000 & n=1 \\
1 &   82257.172 & 10.199 & n=2 \\
2 &  97489.992 & 12.087 & n=3 \\
3 &  102821.219 & 12.748 & n=4 \\
4 &  105288.859  & 13.054 & n=5 \\
5 &  109754.578  & 13.608 & \ion{H}{ii}\\
\end{tabular}
\end{table}

\begin{table}[htpb]
\caption{Calcium energy levels.}
\label{tab:ca}
\begin{tabular}{rrrl}
Level & E (cm$^{-1}$) & E (eV) & Designation\\
\hline
 0 &       0.000 &    0.000 & \ion{Ca}{II} $3p^{6}\,\,4s\,\,^{2}\!S$\\
 1 &   13650.248 &    1.692 & \ion{Ca}{II} $3p^{6}\,\,3d\,\,^{2}\!D_{3/2}$\\
 2 &   13710.900 &    1.700 & \ion{Ca}{II} $3p^{6}\,\,3d\,\,^{2}\!D_{5/2}$\\
 3 &   25191.535 &    3.123 & \ion{Ca}{II} $3p^{6}\,\,4p\,\,^{2}\!P^{o}_{1/2}$\\
 4 &   25414.465 &    3.151 & \ion{Ca}{II} $3p^{6}\,\,4p\,\,^{2}\!P^{o}_{3/2}$\\
 5 &   95751.870 &   11.872 & \ion{Ca}{III} ground\,\,term\\
\end{tabular}
\end{table}

\begin{table}[htpb]
\caption{Helium energy levels.}
\label{tab:he}
\begin{tabular}{rrrl}
Level & E (cm$^{-1}$) & E (eV) & Designation\\
\hline
 0 &       0.000 &    0.000 & \ion{He}{i} $1s^{2}\,\,^{1}\!S_{0}$\\
 1 &  159852.231 &   19.819 & \ion{He}{i} $1s\,\,2s\,\,^{3}\!S_{1}$\\
 2 &  166273.513 &   20.615 & \ion{He}{i} $1s\,\,2s\,\,^{1}\!S_{0}$\\
 3 &  169083.059 &   20.964 & \ion{He}{i} $1s\,\,2p\,\,^{3}\!P^{o}$\\
 4 &  171131.075 &   21.218 & \ion{He}{i} $1s\,\,2p\,\,^{1}\!P^{o}_{1}$\\
 5 &  198420.575 &   24.601 & \ion{He}{ii} $1s\,\,^{2}\!S_{1/2}$\\
 6 &  527600.309 &   65.415 & \ion{He}{ii} $2s\,\,^{2}\!S_{1/2}$\\
 7 &  527603.747 &   65.415 & \ion{He}{ii} $2p\,\,^{2}\!P^{o}$\\
 8 &  637329.434 &   79.019 & \ion{He}{iii}\\
\end{tabular}
\end{table}

\begin{table}[htpb]
\caption{Radiative transitions treated in detail. Wavelengths are given in vacuum ($\lambda < 2000$ \AA)  or air 
($\lambda > 2000$ \AA).}
\label{tab:trans}
\begin{tabular}{rllrll}
\# & Ion & Id & $\lambda$ (\AA) & lower & upper \\
\hline
  0 & \ion{H}{I} & Ly-$\alpha$ &        1215.70 & n=1  $_{}$ & n=2  $_{}$\\
  1 & \ion{H}{I} & Ly-$\beta$ &        1025.75 & n=1  $_{}$ & n=3  $_{}$\\
  2 & \ion{H}{I} & Ly-$\gamma$ &         972.56 & n=1  $_{}$ & n=4  $_{}$\\
  3 & \ion{H}{I} & Ly-$\delta$ &         949.77 & n=1  $_{}$ & n=5  $_{}$\\
  4 & \ion{H}{I} & H-$\alpha$ &        6562.96 & n=2  $_{}$ & n=3  $_{}$\\
  5 & \ion{H}{I} & H-$\beta$ &        4861.50 & n=2  $_{}$ & n=4  $_{}$\\
  6 & \ion{H}{I} & H-$\gamma$ &        4340.62 & n=2  $_{}$ & n=5  $_{}$\\
  7 & \ion{H}{I} & Pa-$\alpha$ &       18752.27 & n=3  $_{}$ & n=4  $_{}$\\
  8 & \ion{H}{I} & Pa-$\beta$ &       12818.86 & n=3  $_{}$ & n=5  $_{}$\\
  9 & \ion{H}{I} & Br-$\alpha$ &       40513.47 & n=4  $_{}$ & n=5  $_{}$\\
 10 & \ion{H}{I} & Ly-cont &         911.12 & n=1  $_{}$ & \ion{H}{ii}\\
 11 & \ion{H}{I} & Ba-cont &        3635.67 & n=2  $_{}$ & \ion{H}{ii}\\
 12 & \ion{H}{I} & Pa-cont &        8151.31 & n=3  $_{}$ & \ion{H}{ii}\\
 13 & \ion{H}{I} & Br-cont &       14419.07 & n=4  $_{}$ & \ion{H}{ii}\\
 14 & \ion{H}{I} & Ph-cont &       22386.68 & n=5  $_{}$ & \ion{H}{ii}\\
 15 & \ion{H}{I} & fake cont &       40000.00 &  & \\
 16 & \ion{Ca}{II} & \ion{Ca}{ii} H &        3968.46 & $3p^{6}\,\,4s\,\,^{2}\!S$ & $3p^{6}\,\,4p\,\,^{2}\!P^{o}_{1/2}$\\
 17 & \ion{Ca}{II} & \ion{Ca}{ii} K &        3933.65 & $3p^{6}\,\,4s\,\,^{2}\!S$ & $3p^{6}\,\,4p\,\,^{2}\!P^{o}_{3/2}$\\
 18 & \ion{Ca}{II} & \ion{Ca}{ii} IR &        8662.16 & $3p^{6}\,\,3d\,\,^{2}\!D_{3/2}$ & $3p^{6}\,\,4p\,\,^{2}\!P^{o}_{1/2}$\\
 19 & \ion{Ca}{II} & \ion{Ca}{ii} IR &        8498.01 & $3p^{6}\,\,3d\,\,^{2}\!D_{3/2}$ & $3p^{6}\,\,4p\,\,^{2}\!P^{o}_{3/2}$\\
 20 & \ion{Ca}{II} & \ion{Ca}{ii} IR &        8542.05 & $3p^{6}\,\,3d\,\,^{2}\!D_{5/2}$ & $3p^{6}\,\,4p\,\,^{2}\!P^{o}_{3/2}$\\
 21 & \ion{Ca}{II} &  &        1044.37 & $3p^{6}\,\,4s\,\,^{2}\!S$ & \ion{Ca}{III} \\
 22 & \ion{Ca}{II} &  &        1218.00 & $3p^{6}\,\,3d\,\,^{2}\!D_{3/2}$ & \ion{Ca}{III} \\
 23 & \ion{Ca}{II} &  &        1218.90 & $3p^{6}\,\,3d\,\,^{2}\!D_{5/2}$ & \ion{Ca}{III} \\
 24 & \ion{Ca}{II} &  &        1417.23 & $3p^{6}\,\,4p\,\,^{2}\!P^{o}_{1/2}$ & \ion{Ca}{III} \\
 25 & \ion{Ca}{II} &  &        1421.72 & $3p^{6}\,\,4p\,\,^{2}\!P^{o}_{3/2}$ & \ion{Ca}{III} \\
 26 & \ion{He}{I} &  &         625.58 & $1s^{2}\,\,^{1}\!S_{0}$ & $1s\,\,2s\,\,^{3}\!S_{1}$\\
 27 & \ion{He}{I} &  &         601.42 & $1s^{2}\,\,^{1}\!S_{0}$ & $1s\,\,2s\,\,^{1}\!S_{0}$\\
 28 & \ion{He}{I} &  &       10830.29 & $1s\,\,2s\,\,^{3}\!S_{1}$ & $1s\,\,2p\,\,^{3}\!P^{o}$\\
 29 & \ion{He}{I} & \ion{He}{i} 584 &         584.35 & $1s^{2}\,\,^{1}\!S_{0}$ & $1s\,\,2p\,\,^{1}\!P^{o}_{1}$\\
 30 & \ion{He}{I} &  &       20580.82 & $1s\,\,2s\,\,^{1}\!S_{0}$ & $1s\,\,2p\,\,^{1}\!P^{o}_{1}$\\
 31 & \ion{He}{II} &  &         303.79 & $1s\,\,^{2}\!S_{1/2}$ & $2s\,\,^{2}\!S_{1/2}$\\
 32 & \ion{He}{II} & \ion{He}{ii} 304 &         303.78 & $1s\,\,^{2}\!S_{1/2}$ & $2p\,\,^{2}\!P^{o}$\\
 33 & \ion{He}{I} &  &         503.98 & $1s^{2}\,\,^{1}\!S_{0}$ & \ion{He}{II} \\
 34 & \ion{He}{I} &  &        2592.02 & $1s\,\,2s\,\,^{3}\!S_{1}$ & \ion{He}{II} \\
 35 & \ion{He}{I} &  &        3109.80 & $1s\,\,2s\,\,^{1}\!S_{0}$ & \ion{He}{II} \\
 36 & \ion{He}{I} &  &        3407.63 & $1s\,\,2p\,\,^{3}\!P^{o}_{4}$ & \ion{He}{II} \\
 37 & \ion{He}{I} &  &        3663.37 & $1s\,\,2p\,\,^{1}\!P^{o}_{1}$ & \ion{He}{II} \\
 38 & \ion{He}{II} &  &         227.84 & $1s\,\,^{2}\!S_{1/2}$ & \ion{He}{III} \\
 39 & \ion{He}{II} &  &         911.34 & $2s\,\,^{2}\!S_{1/2}$ & \ion{He}{III} \\
 40 & \ion{He}{II} &  &         911.36 & $2p\,\,^{2}\!P^{o}_{2}$ & \ion{He}{III} \\
\end{tabular}
\end{table}

\section{Initial atmosphere}\label{sec:atmosphere}

Most \radyn\ simulations of the Sun have started from an atmosphere that is in energy equilibrium, including a mixing-length description of the convective flux divergence in the photosphere, conduction in the corona, and optically thin radiative losses in addition to the detailed radiative transfer in
the transitions in table~\ref{tab:trans}. This starting atmosphere does not have a chromospheric temperature rise --- only a temperature rise above
1~Mm height from the absorption of coronal radiation in the hydrogen Lyman continuum and helium continua. 
Such an initial atmosphere is based on a well-understood energy equation, but any comparison with observations of chromospheric diagnostics suffers from the lack of a chromospheric temperature rise.

Here we present a grid that instead has a starting atmosphere similar to the VAL3C semi-empirical atmosphere 
\citep{1981ApJS...45..635V}. 
We started from the VAL3C atmosphere and extended it with a transition region and a corona, with the temperature fixed at 1~MK at the loop apex. 
The radiative flux divergence per gram of matter was calculated from this model and a term was added to the energy equation to obtain
the energy balance in the photosphere and chromosphere. In the transition region and corona, we wanted the energy balance to be set by the balance between thermal conduction and radiative losses, rather than by the semi-empirical temperature structure in the VAL3C model atmosphere. The extra heating needed to maintain the VAL3C temperature structure is
almost constant (as heating per unit mass) throughout the chromosphere above 1~Mm height. We therefore set this extra heating term to a constant value in the upper atmosphere.
The term was tabulated as a function of
the column mass and kept constant throughout the simulations. 

Before starting the simulation, the atmosphere was relaxed such that the
coronal temperature structure is given by the balance between the conductive flux divergence and the optically thin radiative losses
given the fixed temperature of 1~MK at the loop apex. After this relaxation, a fixed heating was added in the top three zones to
maintain the top temperature in the absence of additional heating to the atmosphere. When beam heating was switched on, the top temperature was thus allowed to increase.

The atmosphere is assumed to be in the form of a quarter-circle loop with a length of 10~Mm. Beyond a distance of 10~Mm along the loop from the apex, the loop is assumed to be vertical down to the lower boundary which is 90~km below the height of $\tau_{\rm 500\,nm}$=1 (which is our zero-point for the z scale). The geometry of the loop only appears in the equations through the 'field-aligned' gravitational force. The boundary at the loop apex is closed --- causing full reflection of any waves. This corresponds to having a semi-circular loop that is symmetric around the apex. The bottom boundary is also closed.

\begin{figure}[htbp]
\includegraphics[width=\columnwidth]{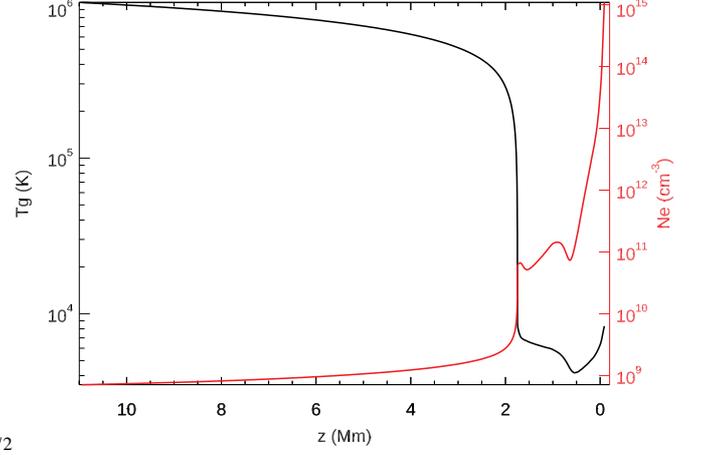}
\caption{Initial atmospheric temperature (black) and electron density (red). The length coordinate z is measured along the loop with zero at $\tau_{\rm 500\,nm}$=1.
The loop apex is at z=11~Mm.
\label{fig:atmos}}
\end{figure}

\begin{figure}[htbp]
\includegraphics[width=\columnwidth]{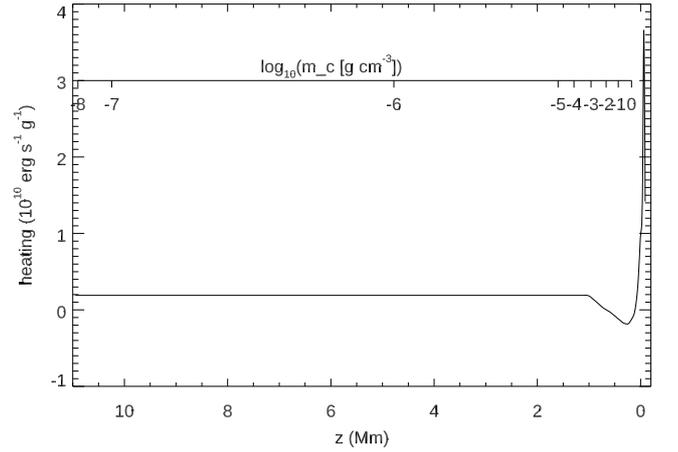}
\caption{Extra heating applied to sustain a VAL3C-like initial atmosphere. The constant value in the
chromosphere, transition region, and corona is $1.9 \times 10^9$ erg~s$^{-1}$ g$^{-1}$.
\label{fig:heatin}}
\end{figure}

The initial temperature structure is shown in Fig~\ref{fig:atmos} with the extra heating term shown in Fig~\ref{fig:heatin}. The equations were
solved on an adaptive grid (see 
\citet{1987JCoPh..69..175D} for a description of the adaptive grid method)
with 300 grid points. In the initial atmosphere, the distance between two grid points is, on average, 37~km with a maximum distance of 57~km and a minimum distance of 0.18~km. During the simulation, the grid adapts to resolve large gradients and the
minimum distance between two grid points may become smaller than 1~m.

The behaviour of the adaptive grid is regulated by weighting factors on the first and second derivatives of the variables. During the construction of the database of models,
some models failed to converge. These were rerun with a different set of adaptive grid weights, typically introducing weights on population densities to better 
resolve ionisation fronts. This means that the models in the F-CHROMA model database do not have identical adaptive grid weights, also giving differences in the positions of the grid points in the initial
atmospheres. All adaptive grid weights are given in the output files.

\section{Model grid}\label{sec:grid}

The F-CHROMA grid of models presented here have the same initial atmosphere (see Section~\ref{sec:atmosphere}); the parameters that vary
are the beam parameters. We employed a triangular shape for the beam energy flux as a function of time with a linear increase in beam flux
from zero at t=0 to a maximum at t=10~s and then a linear decrease to zero beam flux at t=20~s. The rationale for this profile is the typical rapid rise and decay of the impulsive time profiles of the higher energy hard X-ray bursts that map the energy input in the form of non-thermal electrons \citep[e.g.][]{1978SoPh...58..127D,1985SoPh..100..465D}. There is evidence for shorter timescale variations \citep[e.g.][]{1995ApJ...447..923A}, but typically these have a smaller amplitude modulation, and the injection time profile does not include these. The simulations were continued with zero beam flux for an additional 30~s until t=50 s to capture the longer-timescale evolution of the atmosphere once direct beam heating stopped (e.g. the effects of conduction to the chromosphere from the hot coronal loop). The top and bottom boundary conditions were closed so the later time evolution might be influenced by
reflection at the boundaries.
We have varied  the spectral index $\delta$, the total energy, and the low-energy cutoff $E_c$; readers can refer to table~\ref{tab:params} for the values. We note that we label the models with the total energy; a total energy of 
$3\times 10^{10}$ erg~cm$^{-2}$ corresponds to a  maximum energy flux of
$3\times 10^9$ erg~cm$^{-2}$~s$^{-1}$. The output file naming convention is best illustrated with an example:\\
{\tt radyn\_out.val3c\_d3\_1.0e12\_t20s\_25kev\_fp} \\
is the output file with a VAL3C-like starting atmosphere, a spectral index $\delta$=3, a {total} energy of $1\times 10^{12}$ erg~cm$^{-2}$, 
a triangular time variation of the beam flux with a total duration of 20~s, a low-energy cutoff of 25~keV, and a Fokker-Planck description 
of the beam. 

\begin{table}[htbp]
\caption{Parameter values for the grid.}
\label{tab:params}
\begin{tabular}{ll}
Parameter & values \\
\hline
$\delta$ & 3,4,5,6,7,8\\
$E_{\rm tot}$ & $3\times 10^{10}, 1\times 10^{11}, 3\times 10^{11}, 1\times 10^{12}$ erg cm$^{-2}$\\
$E_c$ & 10, 15, 20, 25 keV\\
\end{tabular}
\end{table}

\section{Properties of the grid models}
\begin{figure}[htbp]
\includegraphics[width=\columnwidth]{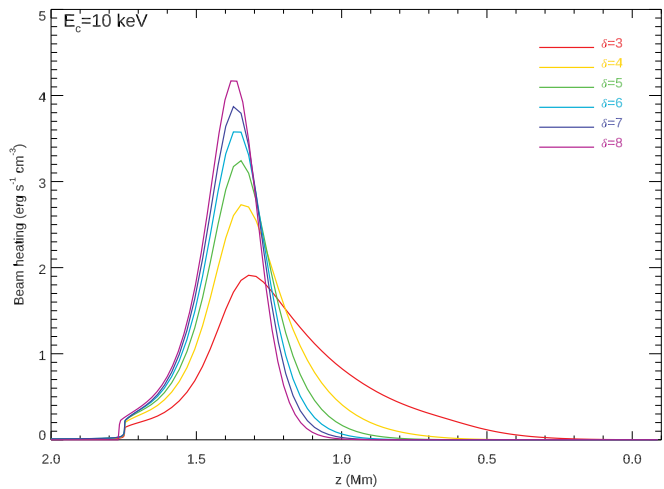}
\caption{Beam heating at t=0.1s as function of height and spectral index for models with a low-energy cutoff of 10~keV. The total beam heating at this time is $1\times 10^{8}$ erg~cm$^{-2}$~s$^{-1}$.
\label{fig:qb_vol_0_1}}
\end{figure}

\begin{figure}[htbp]
\includegraphics[width=\columnwidth]{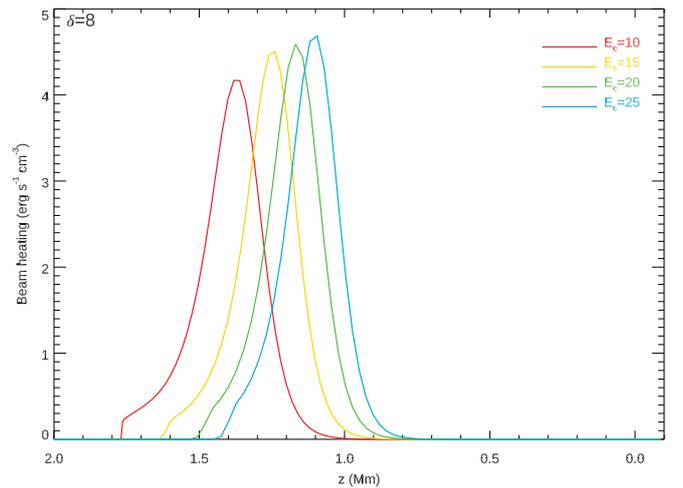}
\caption{Beam heating at t=0.1s as function of height and low-energy cutoff for models with a spectral index of 8. The total beam heating at this time is $1\times 10^{8}$ erg~cm$^{-2}$~s$^{-1}$.
\label{fig:qb_vol_delta_5_1}}
\end{figure}

Different beam parameters lead to differences in the penetration depth of the beam. A lower value of the spectral index $\delta$ leads to a higher proportion of high-energy electrons 
and thus more heating low down in the atmosphere (Fig.~\ref{fig:qb_vol_0_1}). A higher value of the low-energy cutoff, $E_c$, also leads to more high-energy electrons and heating
lower down (Fig.~\ref{fig:qb_vol_delta_5_1}); readers can also refer to the discussion in \citet{2015ApJ...809..104A}.
The figures show heating per volume. Since the density decreases with height because of the hydrostatic stratification, heating per particle decreases deeper in the chromosphere.

\begin{figure*}[htbp]
\includegraphics[width=\textwidth]{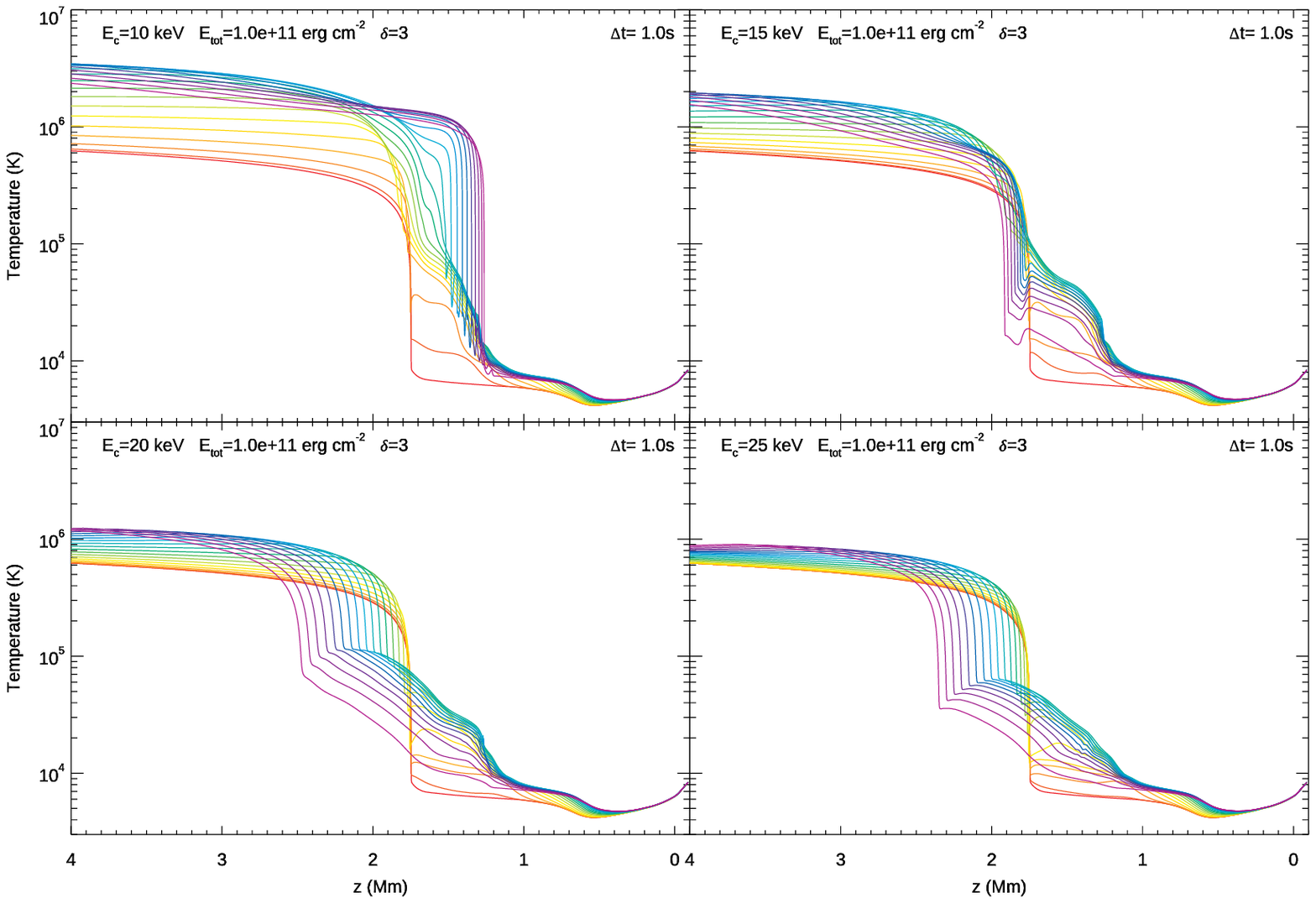}
\caption{Temperature as function of loop position, z, and time for the four different values of the low-energy cutoff, a total energy of $10^{11}$ erg cm$^{-2}$, and a spectral index of 3. The initial temperature structure is shown in red, with subsequent timesteps, separated by 1.0s, shown with rainbow colours from red to violet. The last timestep shown is at t=20s, when the beam was switched off.
\label{fig:fig_fchroma_tgt_z_0_0123_1}}
\end{figure*}

\begin{figure*}[htbp]
\includegraphics[width=\textwidth]{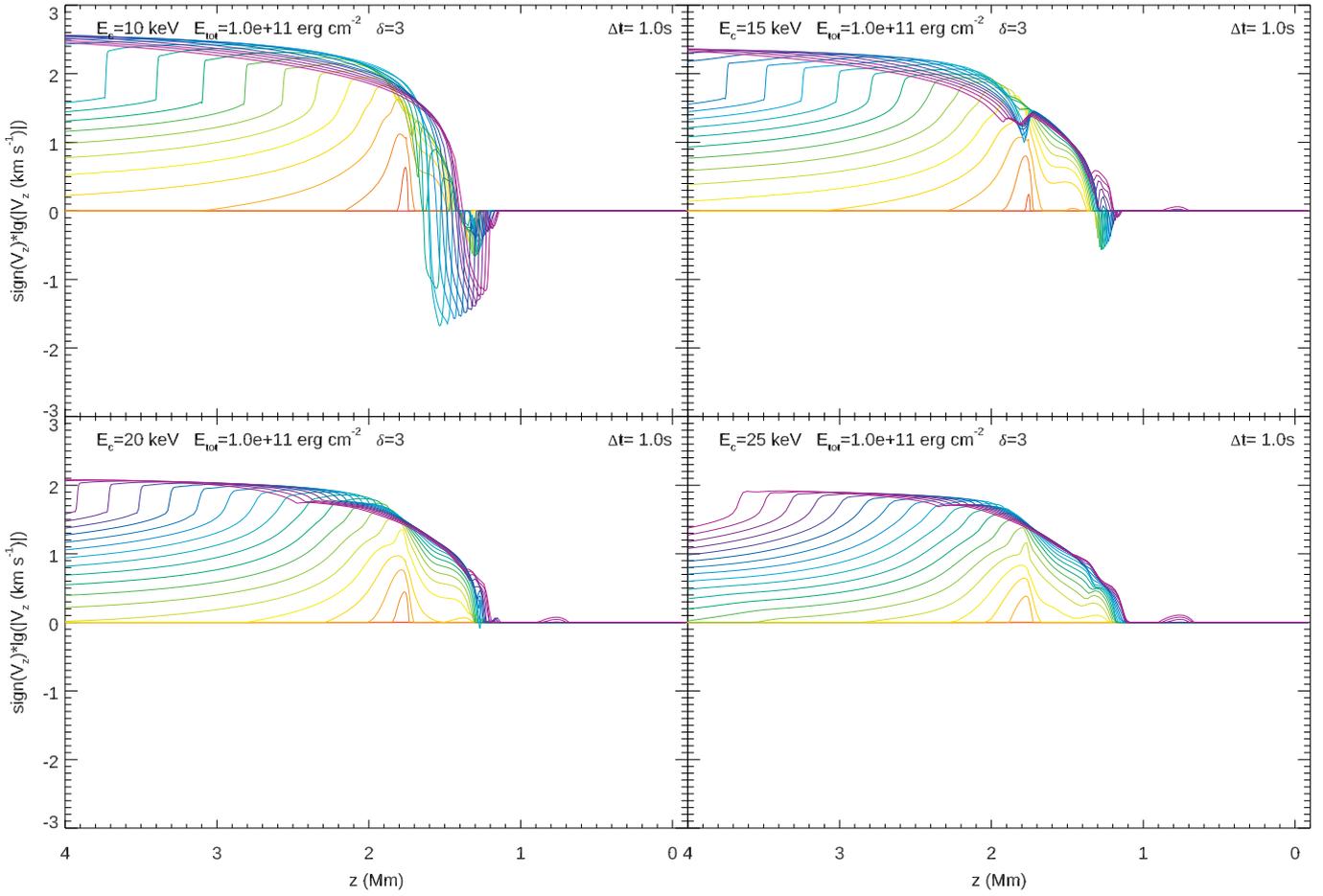}
\caption{Bulk velocity as function of loop position, z, and time for the four different values of the low-energy cutoff, a total energy of $10^{11}$ erg cm$^{-2}$, and a spectral index of 3. The logarithm of the absolute value of the bulk velocity is shown, multiplied by the sign of the velocity. A value of one thus corresponds to an outflow of 10 km~s$^{-1}$, minus one to an inflow of 10 km~s$^{-1}$. Absolute values of the bulk velocity below 1 km~s$^{-1}$ are shown as zero. The initial velocity of zero is shown in red, with subsequent timesteps, separated by 1.0s, shown with rainbow colours from red to violet. The last timestep shown is at t=20s, when the beam was switched off.
\label{fig:fig_fchroma_vzt_z_0_0123_1}}
\end{figure*}

\begin{figure*}[htbp]
\includegraphics[width=\textwidth]{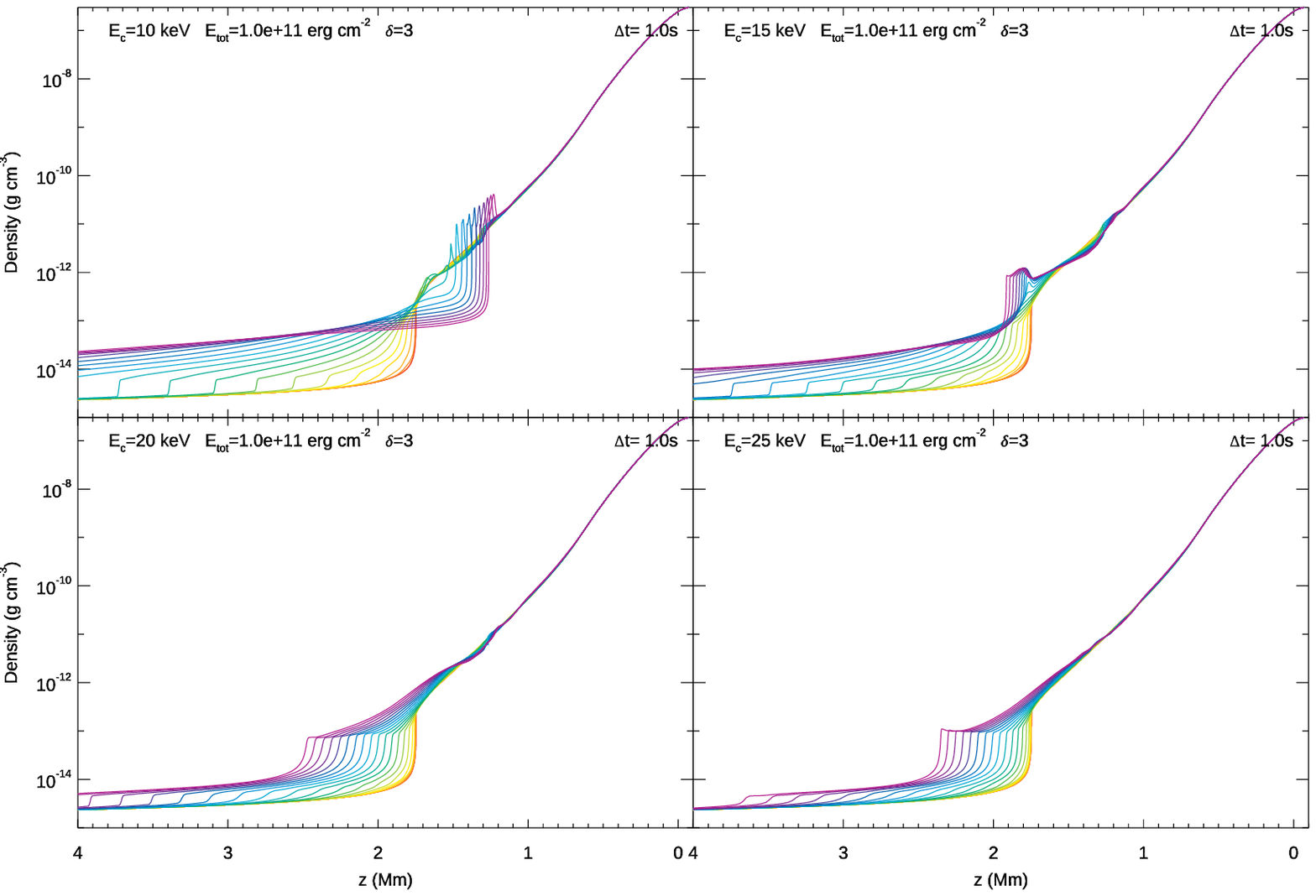}
\caption{Mass density as function of  loop position, z, and time for the four different values of the low-energy cutoff, a total energy of $10^{11}$ erg cm$^{-2}$, and a spectral index of 3. The initial density structure is shown in red, with subsequent timesteps, separated by 1.0s, shown with rainbow colours from red to violet. The last timestep shown is at t=20s, when the beam was switched off.
\label{fig:fig_fchroma_dt_z_0_0123_1}}
\end{figure*}

With time, the heating profile as a function of height changes. The heating causes chromospheric evaporation and ionisation, and the density increases above where the beam heating takes place. This causes more of the beam energy to be deposited at larger heights.
The variation of the heating profile with the beam parameters causes differences in the hydrodynamic response of the atmosphere. Figure \ref{fig:fig_fchroma_tgt_z_0_0123_1} shows the temperature as a function of position, z, and time for the four different values of the low-energy cutoff and a total energy of $10^{11}$ erg cm$^{-2}$ and a spectral index of 3. For the lowest value of the low energy cutoff (upper left panel), the large proportion of low-energy electrons are collisionally stopped in the corona, leading to substantial heating there, expansion of the coronal plasma, and a transition region that moves downwards. A similar phenomenon was observed in simulations by \cite{2009ApJ...702.1553L} and is consistent with hot downflows at flare footpoints such as those that were observed by \cite{2009ApJ...699..968M}. With an increasing low-energy cutoff, the heating moves deeper, and we see less coronal heating, more chromospheric heating, and a transition region moving upwards. This evolution is also seen in the bulk velocities (Fig.\ref{fig:fig_fchroma_vzt_z_0_0123_1}) where the smallest low-energy cutoff leads to large velocities in the corona (outflow of several hundred km~s$^{-1}$) and also large chromospheric condensation velocities, while stopping the beam deeper down at larger densities leads to smaller velocities. The chromospheric condensation leads to an increased density moving downwards and the chromospheric evaporation leads to increased densities higher up (Fig:~\ref{fig:fig_fchroma_dt_z_0_0123_1}).

If the beam energy is deposited in the uppermost part of the chromosphere, we may get a temperature runaway leading to explosive evaporation. This happens because the optically thin loss curve has a maximum around 10$^5\,$K such that when the temperature reaches this value, increased heating leads to decreased radiative losses.
We see this in the temperature evolution with time for different values of the spectral index, $\delta$, in Fig.\ref{fig:fig_fchroma_tgt_z_0345_1_1}. The temperature runaway in the upper chromosphere may leave a chromospheric bubble
between the runaway part and the original transition region; readers can refer to \citet{2020ApJ...894L..21R} for more details.

Obviously, the atmospheric evolution strongly depends on the total energy input. One example of the interplay between coronal and chromospheric heating for the span of total energy is shown in Fig.~\ref{fig:fig_fchroma_tgt_z_0_2_0123}.

These examples are only meant to give an idea of the rich and varied atmospheric behaviour as a function of the beam parameters that can be explored with these models --- an extensive analysis is beyond the scope of this paper. A more detailed analysis of some aspects can be found in the \radyn\ F-CHROMA grid papers referred to in Section~\ref{sec:introduction}.

\section{Grid archive}\label{sec:archive}

All models are available from a searchable wiki-based archive page.%
\footnote{https://star.pst.qub.ac.uk/wiki/doku.php/public/solarmodels/start}
The full RADYN source code is also available.%
\footnote{http://folk.universitetetioslo.no/matsc/radyn/radyn\_fchroma.tar}
Unpacking the tar file creates a directory {\tt radyn\_fchroma} with subdirectories. The file 
{\tt radyn\_fchroma/doc/radyn\_manual.pdf} contains the manual and {\tt radyn\_fchroma/doc/analysis\_tools.pdf} contains the instructions on how to read the CDF files and also a list of variables. 
It is also possible to download just the analysis software and its documentation.%
\footnote{http://folk.universitetetioslo.no/matsc/radyn/tools.tar}

\begin{acknowledgements}
The research leading to these results has received funding from the European Research
Council under the European Union's Seventh Framework Programme (FP7/2007-2013) from
grant n$^o$ 606862 (F-CHROMA) and from 
ERC Grant agreement n$^o$ 291058 (CHROMPHYS).
This research was supported by the Research Council of Norway through its Centres of Excellence scheme, project number 262622, and through grants of computing time from the Programme for Supercomputing.
LF acknowledges support by grants ST/P000533/1 and ST/T000422/1 from the UK's Science and Technology Facilities Council.
PH and JK acknowledge support by grant 19-09489S of the Czech Funding
Agency. They have also been supported by the project RVO:67985815. PJAS acknowledges support from CNPq
(contract 307612/2019-8).

\end{acknowledgements}

\bibliographystyle{aa}
\bibliography{Bibliography}

\begin{thebibliography}{52}
\expandafter\ifx\csname natexlab\endcsname\relax\def\natexlab#1{#1}\fi

\bibitem[{{Abbett} \& {Hawley}(1999)}]{1999ApJ...521..906A}
{Abbett}, W.~P. \& {Hawley}, S.~L. 1999, \apj, 521, 906

\bibitem[{{Allred} {et~al.}(2005){Allred}, {Hawley}, {Abbett}, \&
  {Carlsson}}]{2005ApJ...630..573A}
{Allred}, J.~C., {Hawley}, S.~L., {Abbett}, W.~P., \& {Carlsson}, M. 2005,
  \apj, 630, 573

\bibitem[{{Allred} {et~al.}(2015){Allred}, {Kowalski}, \&
  {Carlsson}}]{2015ApJ...809..104A}
{Allred}, J.~C., {Kowalski}, A.~F., \& {Carlsson}, M. 2015, \apj, 809, 104

\bibitem[{{Arnaud} \& {Rothenflug}(1985)}]{1985A&AS...60..425A}
{Arnaud}, M. \& {Rothenflug}, R. 1985, \aaps, 60, 425

\bibitem[{{Aschwanden} {et~al.}(2016){Aschwanden}, {Holman}, {O'Flannagain},
  {Caspi}, {McTiernan}, \& {Kontar}}]{2016ApJ...832...27A}
{Aschwanden}, M.~J., {Holman}, G., {O'Flannagain}, A., {et~al.} 2016, \apj,
  832, 27

\bibitem[{{Aschwanden} {et~al.}(2019){Aschwanden}, {Kontar}, \&
  {Jeffrey}}]{2019ApJ...881....1A}
{Aschwanden}, M.~J., {Kontar}, E.~P., \& {Jeffrey}, N. L.~S. 2019, \apj, 881, 1

\bibitem[{{Aschwanden} {et~al.}(1995){Aschwanden}, {Schwartz}, \&
  {Alt}}]{1995ApJ...447..923A}
{Aschwanden}, M.~J., {Schwartz}, R.~A., \& {Alt}, D.~M. 1995, \apj, 447, 923

\bibitem[{{Asplund} {et~al.}(2009){Asplund}, {Grevesse}, {Sauval}, \&
  {Scott}}]{2009ARA&A..47..481A}
{Asplund}, M., {Grevesse}, N., {Sauval}, A.~J., \& {Scott}, P. 2009, \araa, 47,
  481

\bibitem[{{Bradshaw} \& {Cargill}(2013)}]{2013ApJ...770...12B}
{Bradshaw}, S.~J. \& {Cargill}, P.~J. 2013, \apj, 770, 12

\bibitem[{{Bradshaw} \& {Mason}(2003)}]{2003A&A...401..699B}
{Bradshaw}, S.~J. \& {Mason}, H.~E. 2003, \aap, 401, 699

\bibitem[{{Capparelli} {et~al.}(2017){Capparelli}, {Zuccarello}, {Romano},
  {Sim{\~o}es}, {Fletcher}, {Kuridze}, {Mathioudakis}, {Keys}, {Cauzzi}, \&
  {Carlsson}}]{2017ApJ...850...36C}
{Capparelli}, V., {Zuccarello}, F., {Romano}, P., {et~al.} 2017, \apj, 850, 36

\bibitem[{{Carlsson} \& {Stein}(1992)}]{1992ApJ...397L..59C}
{Carlsson}, M. \& {Stein}, R.~F. 1992, \apjl, 397, L59

\bibitem[{{Carlsson} \& {Stein}(1995)}]{1995ApJ...440L..29C}
{Carlsson}, M. \& {Stein}, R.~F. 1995, \apjl, 440, L29

\bibitem[{{Carlsson} \& {Stein}(1997)}]{1997ApJ...481..500C}
{Carlsson}, M. \& {Stein}, R.~F. 1997, \apj, 481, 500

\bibitem[{{Carlsson} \& {Stein}(2002)}]{2002ApJ...572..626C}
{Carlsson}, M. \& {Stein}, R.~F. 2002, \apj, 572, 626

\bibitem[{{Cavallini}(2006)}]{2006SoPh..236..415C}
{Cavallini}, F. 2006, \solphys, 236, 415

\bibitem[{{de Jager} \& {de Jonge}(1978)}]{1978SoPh...58..127D}
{de Jager}, C. \& {de Jonge}, G. 1978, \solphys, 58, 127

\bibitem[{{De Pontieu} {et~al.}(2014){De Pontieu}, {Title}, {Lemen}, {Kushner},
  {Akin}, {Allard}, {Berger}, {Boerner}, {Cheung}, {Chou}, {Drake}, {Duncan},
  {Freeland}, {Heyman}, {Hoffman}, {Hurlburt}, {Lindgren}, {Mathur}, {Rehse},
  {Sabolish}, {Seguin}, {Schrijver}, {Tarbell}, {W{\"u}lser}, {Wolfson},
  {Yanari}, {Mudge}, {Nguyen-Phuc}, {Timmons}, {van Bezooijen}, {Weingrod},
  {Brookner}, {Butcher}, {Dougherty}, {Eder}, {Knagenhjelm}, {Larsen},
  {Mansir}, {Phan}, {Boyle}, {Cheimets}, {DeLuca}, {Golub}, {Gates}, {Hertz},
  {McKillop}, {Park}, {Perry}, {Podgorski}, {Reeves}, {Saar}, {Testa}, {Tian},
  {Weber}, {Dunn}, {Eccles}, {Jaeggli}, {Kankelborg}, {Mashburn}, {Pust},
  {Springer}, {Carvalho}, {Kleint}, {Marmie}, {Mazmanian}, {Pereira}, {Sawyer},
  {Strong}, {Worden}, {Carlsson}, {Hansteen}, {Leenaarts}, {Wiesmann},
  {Aloise}, {Chu}, {Bush}, {Scherrer}, {Brekke}, {Martinez-Sykora}, {Lites},
  {McIntosh}, {Uitenbroek}, {Okamoto}, {Gummin}, {Auker}, {Jerram}, {Pool}, \&
  {Waltham}}]{2014SoPh..289.2733D}
{De Pontieu}, B., {Title}, A.~M., {Lemen}, J.~R., {et~al.} 2014, \solphys, 289,
  2733

\bibitem[{{Dennis}(1985)}]{1985SoPh..100..465D}
{Dennis}, B.~R. 1985, \solphys, 100, 465

\bibitem[{{Dere} {et~al.}(1997){Dere}, {Landi}, {Mason}, {Monsignori Fossi}, \&
  {Young}}]{1997A&AS..125..149D}
{Dere}, K.~P., {Landi}, E., {Mason}, H.~E., {Monsignori Fossi}, B.~C., \&
  {Young}, P.~R. 1997, \aaps, 125, 149

\bibitem[{{Dorfi} \& {Drury}(1987)}]{1987JCoPh..69..175D}
{Dorfi}, E.~A. \& {Drury}, L.~O. 1987, Journal of Computational Physics, 69,
  175

\bibitem[{{Druett} \& {Zharkova}(2019)}]{2019A&A...623A..20D}
{Druett}, M.~K. \& {Zharkova}, V.~V. 2019, \aap, 623, A20

\bibitem[{{Fang} {et~al.}(1993){Fang}, {Henoux}, \&
  {Gan}}]{1993A&A...274..917F}
{Fang}, C., {Henoux}, J.~C., \& {Gan}, W.~Q. 1993, \aap, 274, 917

\bibitem[{{Hannah} {et~al.}(2011){Hannah}, {Hudson}, {Battaglia}, {Christe},
  {Ka{\v{s}}parov{\'a}}, {Krucker}, {Kundu}, \&
  {Veronig}}]{2011SSRv..159..263H}
{Hannah}, I.~G., {Hudson}, H.~S., {Battaglia}, M., {et~al.} 2011, \ssr, 159,
  263

\bibitem[{{Heinzel} {et~al.}(2016){Heinzel}, {Ka{\v{s}}parov{\'a}}, {Varady},
  {Karlick{\'y}}, \& {Moravec}}]{2016IAUS..320..233H}
{Heinzel}, P., {Ka{\v{s}}parov{\'a}}, J., {Varady}, M., {Karlick{\'y}}, M., \&
  {Moravec}, Z. 2016, in Solar and Stellar Flares and their Effects on Planets,
  ed. A.~G. {Kosovichev}, S.~L. {Hawley}, \& P.~{Heinzel}, Vol. 320, 233--238

\bibitem[{{Huang} {et~al.}(2020){Huang}, {Sadykov}, {Xu}, {Jing}, \&
  {Wang}}]{2020ApJ...897L...6H}
{Huang}, N., {Sadykov}, V.~M., {Xu}, Y., {Jing}, J., \& {Wang}, H. 2020, \apjl,
  897, L6

\bibitem[{{Jeffrey} {et~al.}(2018){Jeffrey}, {Fletcher}, {Labrosse}, \&
  {Sim{\~o}es}}]{2018SciA....4.2794J}
{Jeffrey}, N.~L.~S., {Fletcher}, L., {Labrosse}, N., \& {Sim{\~o}es}, P.~J.~A.
  2018, Science Advances, 4, 2794

\bibitem[{{Jess} {et~al.}(2010){Jess}, {Mathioudakis}, {Christian}, {Keenan},
  {Ryans}, \& {Crockett}}]{2010SoPh..261..363J}
{Jess}, D.~B., {Mathioudakis}, M., {Christian}, D.~J., {et~al.} 2010, \solphys,
  261, 363

\bibitem[{{Ka{\v s}parov{\'a}} {et~al.}(2009){Ka{\v s}parov{\'a}}, {Varady},
  {Heinzel}, {Karlick{\'y}}, \& {Moravec}}]{2009A&A...499..923K}
{Ka{\v s}parov{\'a}}, J., {Varady}, M., {Heinzel}, P., {Karlick{\'y}}, M., \&
  {Moravec}, Z. 2009, \aap, 499, 923

\bibitem[{{Kowalski} {et~al.}(2017){Kowalski}, {Allred}, {Daw}, {Cauzzi}, \&
  {Carlsson}}]{2017ApJ...836...12K}
{Kowalski}, A.~F., {Allred}, J.~C., {Daw}, A., {Cauzzi}, G., \& {Carlsson}, M.
  2017, \apj, 836, 12

\bibitem[{{Kuridze} {et~al.}(2015){Kuridze}, {Mathioudakis}, {Sim{\~o}es},
  {Rouppe van der Voort}, {Carlsson}, {Jafarzadeh}, {Allred}, {Kowalski},
  {Kennedy}, {Fletcher}, {Graham}, \& {Keenan}}]{2015ApJ...813..125K}
{Kuridze}, D., {Mathioudakis}, M., {Sim{\~o}es}, P.~J.~A., {et~al.} 2015, \apj,
  813, 125

\bibitem[{{Landi} {et~al.}(2012){Landi}, {Del Zanna}, {Young}, {Dere}, \&
  {Mason}}]{2012ApJ...744...99L}
{Landi}, E., {Del Zanna}, G., {Young}, P.~R., {Dere}, K.~P., \& {Mason}, H.~E.
  2012, \apj, 744, 99

\bibitem[{{Leenaarts} {et~al.}(2012){Leenaarts}, {Carlsson}, \& {Rouppe van der
  Voort}}]{2012ApJ...749..136L}
{Leenaarts}, J., {Carlsson}, M., \& {Rouppe van der Voort}, L. 2012, \apj, 749,
  136

\bibitem[{{Liu} {et~al.}(2009){Liu}, {Petrosian}, \&
  {Mariska}}]{2009ApJ...702.1553L}
{Liu}, W., {Petrosian}, V., \& {Mariska}, J.~T. 2009, \apj, 702, 1553

\bibitem[{{Machado} {et~al.}(1980){Machado}, {Avrett}, {Vernazza}, \&
  {Noyes}}]{1980ApJ...242..336M}
{Machado}, M.~E., {Avrett}, E.~H., {Vernazza}, J.~E., \& {Noyes}, R.~W. 1980,
  \apj, 242, 336

\bibitem[{{Milligan} \& {Dennis}(2009)}]{2009ApJ...699..968M}
{Milligan}, R.~O. \& {Dennis}, B.~R. 2009, \apj, 699, 968

\bibitem[{{Monson} {et~al.}(2021){Monson}, {Mathioudakis}, {Reid}, {Milligan},
  \& {Kuridze}}]{2021ApJ...915...16M}
{Monson}, A.~J., {Mathioudakis}, M., {Reid}, A., {Milligan}, R., \& {Kuridze},
  D. 2021, \apj, 915, 16

\bibitem[{{Morgachev} {et~al.}(2021){Morgachev}, {Tsap}, {Motorina},
  {Smirnova}, \& {Motorin}}]{2021Ge&Ae..61.1045M}
{Morgachev}, A.~S., {Tsap}, Y.~T., {Motorina}, G.~G., {Smirnova}, V.~V., \&
  {Motorin}, A.~S. 2021, Geomagnetism and Aeronomy, 61, 1045

\bibitem[{{Morgachev} {et~al.}(2020){Morgachev}, {Tsap}, {Smirnova},
  {Motorina}, \& {B{\'a}rta}}]{2020Ge&Ae..60.1038M}
{Morgachev}, A.~S., {Tsap}, Y.~T., {Smirnova}, V.~V., {Motorina}, G.~G., \&
  {B{\'a}rta}, M. 2020, Geomagnetism and Aeronomy, 60, 1038

\bibitem[{{Osborne} {et~al.}(2019){Osborne}, {Armstrong}, \&
  {Fletcher}}]{2019ApJ...873..128O}
{Osborne}, C. M.~J., {Armstrong}, J.~A., \& {Fletcher}, L. 2019, \apj, 873, 128

\bibitem[{{Osborne} \& {Mili{\'c}}(2021)}]{2021ApJ...917...14O}
{Osborne}, C. M.~J. \& {Mili{\'c}}, I. 2021, \apj, 917, 14

\bibitem[{{Reep} {et~al.}(2019){Reep}, {Bradshaw}, {Crump}, \&
  {Warren}}]{2019ApJ...871...18R}
{Reep}, J.~W., {Bradshaw}, S.~J., {Crump}, N.~A., \& {Warren}, H.~P. 2019,
  \apj, 871, 18

\bibitem[{{Reid} {et~al.}(2020){Reid}, {Zhigulin}, {Carlsson}, \&
  {Mathioudakis}}]{2020ApJ...894L..21R}
{Reid}, A., {Zhigulin}, B., {Carlsson}, M., \& {Mathioudakis}, M. 2020, \apjl,
  894, L21

\bibitem[{{Sadykov} {et~al.}(2020){Sadykov}, {Kosovichev}, {Kitiashvili}, \&
  {Kerr}}]{2020ApJ...893...24S}
{Sadykov}, V.~M., {Kosovichev}, A.~G., {Kitiashvili}, I.~N., \& {Kerr}, G.~S.
  2020, \apj, 893, 24

\bibitem[{{Sadykov} {et~al.}(2019){Sadykov}, {Kosovichev}, {Sharykin}, \&
  {Kerr}}]{2019ApJ...871....2S}
{Sadykov}, V.~M., {Kosovichev}, A.~G., {Sharykin}, I.~N., \& {Kerr}, G.~S.
  2019, \apj, 871, 2

\bibitem[{{Sim{\~o}es} {et~al.}(2017){Sim{\~o}es}, {Kerr}, {Fletcher},
  {Hudson}, {Gim{\'e}nez de Castro}, \& {Penn}}]{2017A&A...605A.125S}
{Sim{\~o}es}, P.~J.~A., {Kerr}, G.~S., {Fletcher}, L., {et~al.} 2017, \aap,
  605, A125

\bibitem[{{Uitenbroek}(2001)}]{2001ApJ...557..389U}
{Uitenbroek}, H. 2001, \apj, 557, 389

\bibitem[{{Uitenbroek}(2002)}]{2002ApJ...565.1312U}
{Uitenbroek}, H. 2002, \apj, 565, 1312

\bibitem[{{Varady} {et~al.}(2010){Varady}, {Kasparova}, {Moravec}, {Heinzel},
  \& {Karlicky}}]{2010ITPS...38.2249V}
{Varady}, M., {Kasparova}, J., {Moravec}, Z., {Heinzel}, P., \& {Karlicky}, M.
  2010, IEEE Transactions on Plasma Science, 38, 2249

\bibitem[{{Vernazza} {et~al.}(1981){Vernazza}, {Avrett}, \&
  {Loeser}}]{1981ApJS...45..635V}
{Vernazza}, J.~E., {Avrett}, E.~H., \& {Loeser}, R. 1981, \apjs, 45, 635

\bibitem[{{Warmuth} \& {Mann}(2013)}]{2013A&A...552A..86W}
{Warmuth}, A. \& {Mann}, G. 2013, \aap, 552, A86

\bibitem[{{Younger}(1981)}]{1981JQSRT..26..329Y}
{Younger}, S.~M. 1981, \jqsrt, 26, 329

\end{thebibliography}
\FloatBarrier

\begin{appendix}
\section{Additional figures}

\begin{figure*}[htbp]
\includegraphics[width=\textwidth]{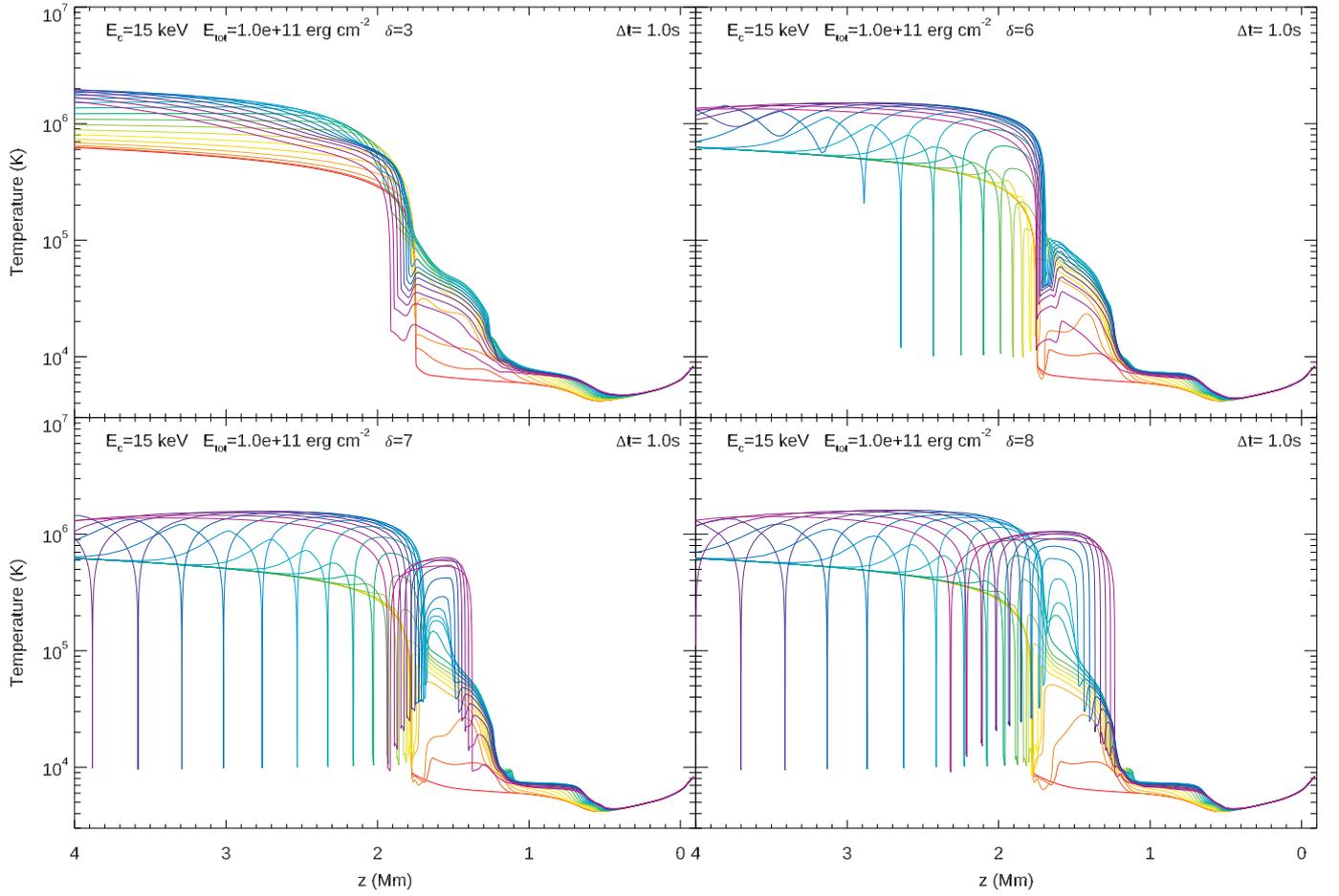}
\caption{Temperature as function of loop position, z, and time for four different values of the spectral index, $\delta$, a lower-energy cutoff of 15~keV, and a total energy of $10^{11}$ erg cm$^{-2}$. The initial temperature structure is shown in red, with subsequent timesteps, separated by 1.0s, shown with rainbow colours from red to violet. The last timestep shown is at t=20s, when the beam was switched off.
\label{fig:fig_fchroma_tgt_z_0345_1_1}}
\end{figure*}

\begin{figure*}[htbp]
\includegraphics[width=\textwidth]{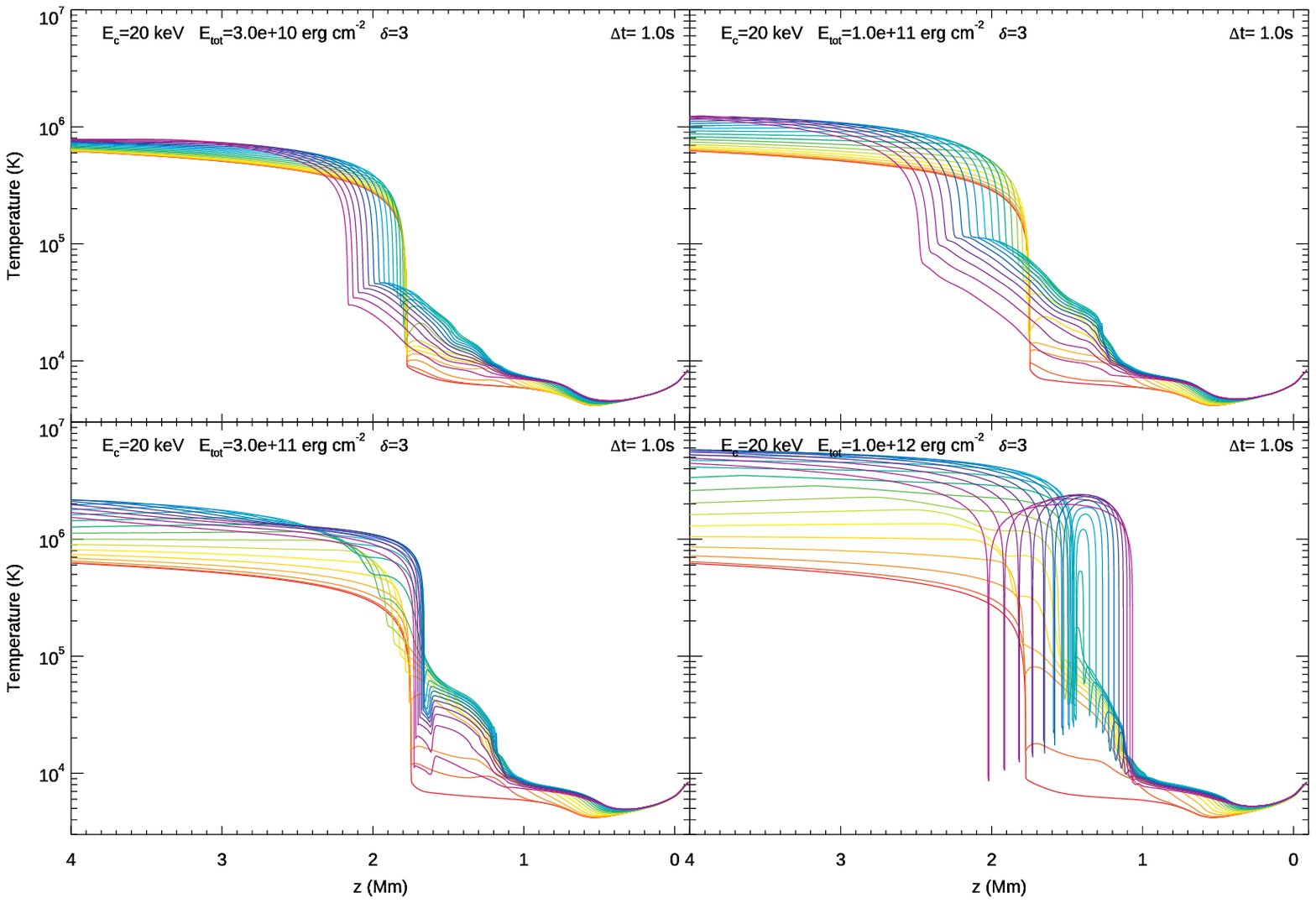}
\caption{Temperature as function of loop position, z, and time for the four different values of the total energy, a lower-energy cutoff of 20~keV, and a spectral index of 3. The initial temperature structure is shown in red, with subsequent timesteps, separated by 1.0s, shown with rainbow colours from red to violet. The last timestep shown is at t=20s, when the beam was switched off.
\label{fig:fig_fchroma_tgt_z_0_2_0123}}
\end{figure*}
\end{appendix}

\end{document}